\documentclass[aip,jcp,reprint,letterpaper,onecolumn,longbibliography]{revtex4-1}
\usepackage{amssymb}
\usepackage{amsmath}
\usepackage[usenames,dvipsnames]{xcolor}
\usepackage{graphicx}
\usepackage{float}
\usepackage[normalem]{ulem}
\usepackage[top=1 in, bottom=1 in]{geometry}
\usepackage{natbib}

\usepackage{xcolor}
\begin{document}

\title
{Delocalization errors in density functional theory are essentially quadratic \textcolor{black}{in fractional occupation number}}
\author{Diptarka Hait}
\affiliation
{{Kenneth S. Pitzer Center for Theoretical Chemistry, Department of Chemistry, University of California, Berkeley, California 94720, USA}}

\author{Martin Head-Gordon}
\email{mhg@cchem.berkeley.edu}
\affiliation
{{Kenneth S. Pitzer Center for Theoretical Chemistry, Department of Chemistry, University of California, Berkeley, California 94720, USA}}
\affiliation{Chemical Sciences Division, Lawrence Berkeley National Laboratory, Berkeley, California 94720, USA}
	
\begin{abstract}
Approximate functionals used in practical density functional theory (DFT) deviate from the piecewise linear behavior of the exact functional for fractional charges. This deviation causes excess charge delocalization, which leads to incorrect densities, molecular properties, barrier heights, band gaps and excitation energies. We present a simple delocalization function for characterizing this error and find it to be almost perfectly linear vs the fractional electron number for systems spanning in size from the H atom to the C$_{12}$H$_{14}$ polyene. This causes the delocalization energy error to be \textcolor{black}{a} quadratic \textcolor{black}{polynomial} in the fractional electron number, which permits us to assess the comparative performance of 47 popular and recent functionals through the curvature. The quadratic form further suggests that information about a single fractional charge is sufficient to eliminate the principal source of delocalization error. Generalizing traditional two-point information like ionization potentials or electron affinities to account for a third, fractional charge based data point could therefore permit fitting/tuning of functionals with lower delocalization error.   
\end{abstract}
	\maketitle
The central idea behind density functional theory (DFT) is that there exists a functional which maps the ground state electron density of an electron distribution to its energy\cite{hohenberg1964inhomogeneous,kohn1965self,becke2014perspective,jones2015density,mardirossian2017thirty}. The existence of this exact functional was formally proven by Hohenberg and Kohn in 1964\cite{hohenberg1964inhomogeneous}, but it remains computationally inaccessible for realistic systems. \textcolor{black}{It is however possible to express the total energy $E$ as : 
\begin{align}
E&=E_T+E_{ext}+E_J+E_{xc}\label{DFT_energy}
\end{align}
where the non-interacting kinetic energy $E_T$, the electron-external potential interaction energy $E_{ext}$ (which includes electron-nuclear attraction energy) and the quasi-classical electron repulsion $E_J$ are exactly known, leaving behind the exchange correlation energy $E_{xc}$ as the only unknown term}\cite{kohn1965self}.
The practical use of DFT typically involves employing some of the hundreds of density functional approximations (DFAs) \textcolor{black}{for $E_{xc}$
} that have been proposed over the last few decades\cite{mardirossian2017thirty}. 
DFAs deviate from the exact functional in many (at times, predictable) regards and there is often substantial variation in predictions from different DFAs. Despite these shortcomings, DFT has found wide use in chemistry, physics and material science as an often adequately accurate and computationally efficient theory that permits exploration of systems well beyond the reach of more expensive wave function approaches. 

\begin{figure}[htb!]
	\centering
	\begin{minipage}{0.48\textwidth}
		\includegraphics[width=\linewidth]{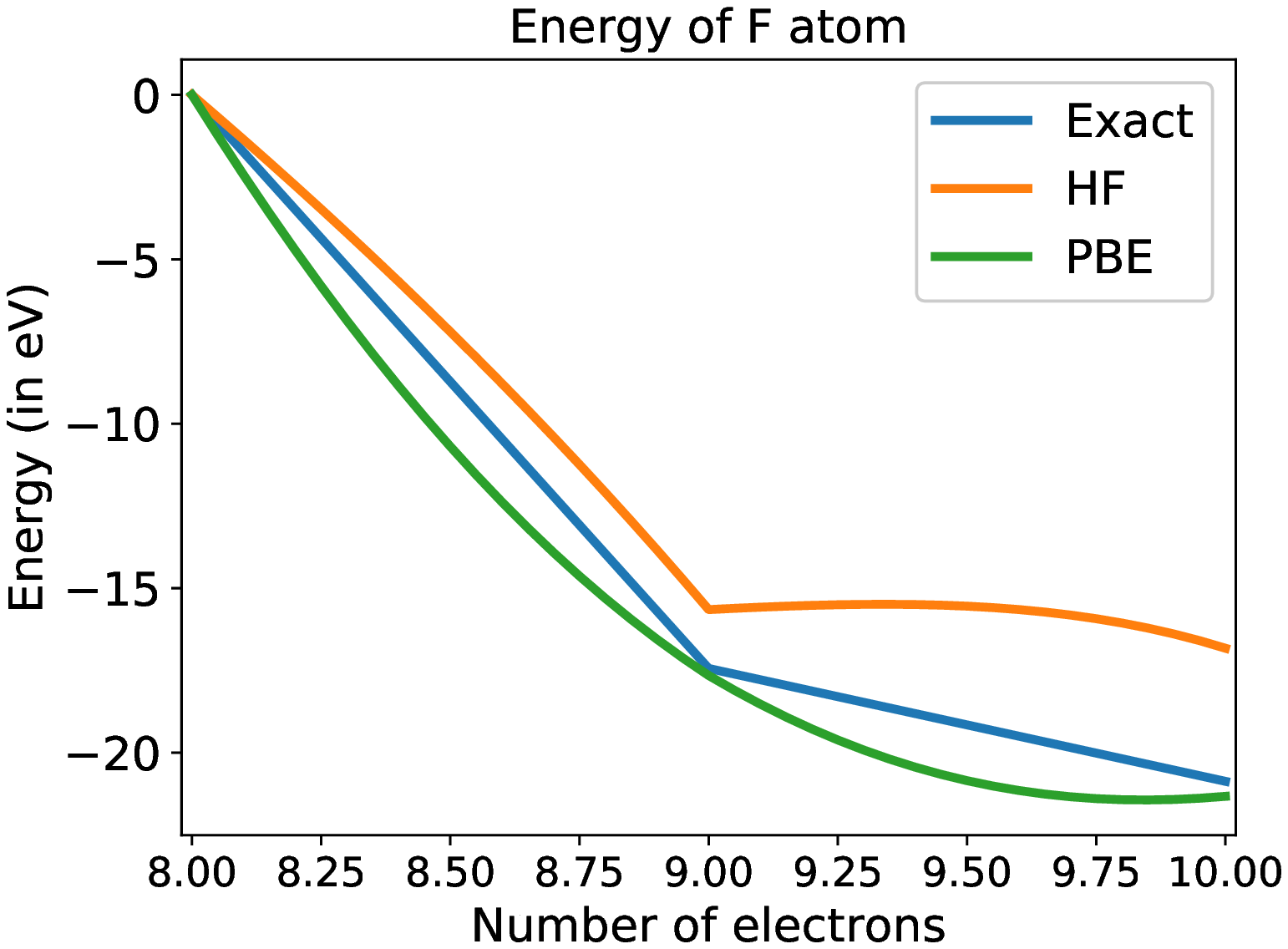}
	\end{minipage}
	\begin{minipage}{0.48\textwidth}
		\includegraphics[width=\linewidth]{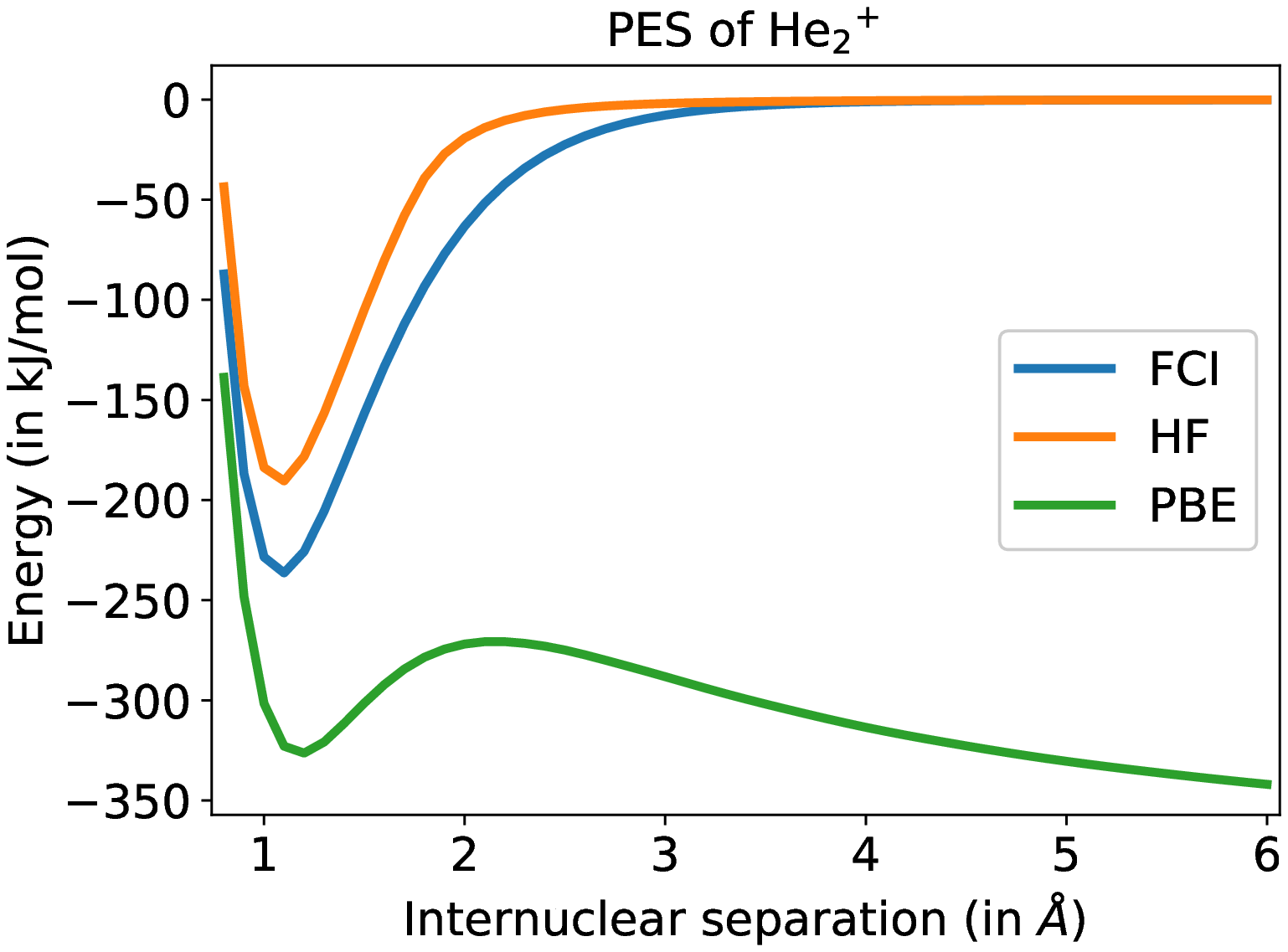}
	\end{minipage}
	\caption{Effect of delocalization error on fractional charges (left) and dissociation of charged species (right), showing overestimation of energy by HF and underestimation by a typical functional (PBE\cite{PBE}). The reported energies are relative to the \textcolor{black}{F$^+$ cation} on the left panel and the true dissociation limit of the He$_2^+$ complex (He$^+$+He) on the right panel, as predicted by each method. FCI on the right panel stands for full configuration interaction. `Exact' behavior on the left panel is obtained from linear interpolation between CCSD(T)\cite{raghavachari1989fifth}/CBS energies. \textcolor{black}{The aug-pc-4 basis\cite{jensen2001polarization,jensen2002polarization,jensen2002polarizationiii,jensen2004polarization,jensen2007polarization} was employed for HF and PBE on the left panel, while all calculations on the right panel used the aug-cc-pVTZ basis\cite{woon1995gaussian}. The basis sets therefore contain reasonable numbers of valence and polarization functions, but are quite incomplete with respect to diffuse functions.}}
	\label{fig:delocerror}
\end{figure}

There are however certain known properties of the exact functional that most modern DFAs deviate from. Using statistical mixture, Perdew et.al.\cite{perdew1982density} showed that the electronic energy of a system with fractional charge is exactly determined by a linear interpolation between the energies corresponding to the two closest integer electron numbers. Mathematically, therefore, the energy $E$ of a system with $N-x$ electrons (where $N$ is a nonnegative integer and $x$ lies between $0$ and $1$) is given by:
\begin{align}
E(N-x)&=E(N)+x(E(N-1)-E(N))\label{linear}
\end{align}
Eqn. \ref{linear} therefore specifies that the electronic energy is piecewise linear with respect to the electron number. It does not specify the energies for integer electron numbers themselves, but difference between molecular electron affinity (EA), $E(N+1)-E(N)$, and ionization potential (IP) $E(N-1)-E(N)$, means that a derivative discontinuity in the energy as a function of electron number occurs at $N$\cite{perdew1982density}. This is illustrated by the exact curve for the F atom, on the left-hand panel of Fig. \ref{fig:delocerror}.
Several other proofs of piecewise linearity have been given \textcolor{black}{elsewhere}\cite{yang2000degenerate,ayers2008dependence,ayers2017levy}.

A particularly simple proof is that of Yang et.al.\cite{yang2000degenerate} who constructed a supersystem of $q$ identical non-interacting copies of an $N$-electron molecule, from which $a<b$ electrons are removed. From one perspective the supersystem is therefore $p$ copies of $E(N-1)$ and $q-p$ copies of $E(N)$. Equivalently, as mixing degenerate states does not affect the energy, the supersystem can be composed of $b$ identical fragments, each with a fractional electron $x=p/q$ removed, with total energy that is evidently:
\begin{align}
q E\left( {N - x} \right) = \left( {q - p} \right)E\left( N \right) + pE\left( {N - 1} \right)
\end{align}
Dividing through by $q$ yields Eqn. \ref{linear} for rational $x$. This was generalized to irrational $x$ by Ayers \cite{ayers2008dependence}. 

Most approximate functionals fail to satisfy Eqn \ref{linear}, giving energies for fractional electron numbers that are typically too low\cite{perdew1985kohn,zhang1998challenge} (as can be seen on the left panel of Fig \ref{fig:delocerror}). This error is often explained in terms of electron self-interaction. \textcolor{black}{The classical electrostatic electron repulsion in Eqn. \ref{DFT_energy} for a density $n(\vec{r})$ is $E_J = \dfrac{1}{2}\displaystyle\int d\vec{r}_1\displaystyle\int \dfrac{n(\vec{r}_1)n(\vec{r}_2)}{|\vec{r}_1-\vec{r}_2|}d\vec{r}_2$}. 
However, $n(\vec{r})$ is made up of a discrete number of electrons, which (despite being delocalized over space) \textit{should not be able to interact with themselves}. \textcolor{black}{$E_J$ therefore contains a spurious self-repulsion energy}. The exchange-correlation energy ($E_{xc}$) in principle should fix this, but the correction is very approximate for most functionals on account of the $E_{xc}$ functional having substantial local components, while self-repulsion is explicitly nonlocal. The electron density subsequently delocalizes (becomes less compact) in order to minimize the residual self-repulsion, and this tends to generate fractional charges whose energies are lower than what Eqn \ref{linear} predicts. This general overdelocalization behavior leads to catastrophes like incorrect asymptotics in dissociation curves for charged species\cite{grafenstein2004impact} (as shown in the right panel of Fig \ref{fig:delocerror}), fractional electrons dissociating from anions\cite{vydrov2007tests} or spurious fractional charges at the dissociation limit for polar bonds\cite{Dutoi2006,Ruzsinszky2006,hait2018accurate,hait2018communication}. It also causes other complications like lowering of barrier heights\cite{patchkovskii2002improving,mori2006many,lynch2000adiabatic}, reduction of band gaps\cite{cohen2008fractional,mori2008localization} and dramatic underestimation of excited state energies with linear response time dependent density functional theory (TDDFT)\cite{dreuw2005single,dreuw2003long,hait2016prediction,bredas}. 

Self-repulsion alone is however an inadequate explanation for deviation from Eqn \ref{linear}\cite{mori2006many,ruzsinszky2007density}. Hartree-Fock (HF) theory lacks electron self-repulsion, but nonetheless does not adhere to Eqn \ref{linear} aside from the trivial single electron cases. In fact, HF predicts \textit{too high} an energy for fractional charges on account of missing correlation energy (as can be seen \textcolor{black}{from the concave HF curves} on the left panel of Fig \ref{fig:delocerror}), preventing HF inspired solutions to self-repulsion (like the Perdew-Zunger correction\cite{pz81}) from satisfying Eqn \ref{linear} for general systems. This effect is often termed as `many electron self-interaction error'\cite{mori2006many}, though it is somewhat misleading as electrons are not really interacting with themselves in this case! We will therefore use the term `self-interaction error' to only refer to incorrect energy predictions arising from an electron spuriously repelling itself, and reserve the more general term `delocalization error' for any and all deviations from Eqn \ref{linear}. Delocalization error in HF manifests itself through overlocalization of electron density, leading to incorrect behavior such as too high barrier heights\cite{lynch2000adiabatic} and too quick a decay in dipole moments during bond dissociation\cite{hait2018accurate}. Effects of this can be seen on the left panel of Fig \ref{fig:delocerror}, where HF predicts too high an energy for systems with fractional charges due to missing correlation energy. 

A study of fractional electron systems offers a formally exact route to characterizing delocalization error and help potentially address it to some extent. Several studies have examined this aspect\cite{vydrov2007tests,mori2006many,cohen2008fractional,mussard2017fractional,lundberg2005quantifying,zhao2016global,bajaj2017communication,dabo2010koopmans} and the rCAM-B3LYP\cite{rcamb3lyp} functional was fitted to fractionally charged carbon atoms. Naturally, these studies mostly focused on the energy errors predicted by functionals, and do not appear to have investigated the precise mathematical nature of delocalization error (i.e. its functional form). In this work, we attempt to bridge this gap by studying the fractional occupation behavior of small to moderate molecular systems with 47 popular and recent DFAs\cite{Slater,PW92,PBE,b88,lyp,b3lyp,b97d,b97,b97-2,b97d,b97mv,bmk,verma2014increasing,jin2016qtp,hsehjs,lrcwpbe,lrcwpbeh,m05,m06,m06hf,m06l,m11l,m11,mn12l,mn15l,n12,n12sx,MN15,ms2,pbe0,pw6b95,scan0,SCAN,sogga11x,sogga11,tpss,tpssh,camb3lyp,wellendorff2014mbeef,revm06l,rcamb3lyp,wB97MV,wB97XD,wM05D,wb97xv} from the first four rungs of Jacob's ladder\cite{perdew2001jacob}. The computational approach is to remove $x$ electrons from each molecule (see Table \ref{atoms}), and self-consistently converge the Kohn-Sham equations for that fractional occupation number. For $x=p/q$, this is numerically identical to a calculation on $q$ non-interacting copies of the molecule with $p$ electrons removed. The result for each functional is an $E(N-x)$ curve which can be compared against the straight line connecting calculations with the same functional for $E(N)$ and $E(N-1)$. 

For a general electronic structure method, we can define $f(x)$ such that: 
\begin{align}
E(N-x)&=E(N)+xf(x)(E(N-1)-E(N))\label{gen}
\end{align}
This trivially implies that $f(x=1)=1$ always, though it can deviate from the ideal value of $f(x)=1$ (which leads to Eqn \ref{linear}) for other $x$. $f(x)$ therefore can be termed as a delocalization function, \textcolor{black}{as $f(x)-1$} measures extent of deviation from Eqn. \ref{linear}.
\textcolor{black}{Since $x \in \left( {0,1} \right)$, $f(x)-1$ ought to be representable by a power series in $x$, although there is no guarantee that a finite truncation of the series would serve as an effective approximation.}


\begin{table}[htb!]
	\begin{tabular}{ll|ll}
		Species & $N$ & Species & $N$  \\ \hline 
		H       & 1 & NH$_3$     & 10  \\
		He      & 2 & CN$^-$      & 14 \\
		F$^-$       & 10 & H$_2$O     & 10 
	\end{tabular}
	\caption{Small species studied near the basis set limit \textcolor{black}{(with respect to valence and polarization functions)}, with the larger integer number of electrons used. }
		\label{atoms}
\end{table}

We initially studied the behavior of $f(x)$ for the small systems listed in Table \ref{atoms} with \textcolor{black}{the aug-pc-4 basis\cite{jensen2001polarization,jensen2002polarization,jensen2002polarizationiii,jensen2004polarization,jensen2007polarization}, which should be near the basis set limit with respect to valence and polarization functions (although not diffuse functions--see Computational Methods for details)}. To our surprise, we found that $f(x)$ is almost always well described by a linear fit of the form $ax+b$ , leading to:
\begin{align}
E(N-x)&\approx E(N)+x(ax+b)(E(N-1)-E(N))\label{quad}
\end{align}
\textcolor{black}{It is important to note that all our calculations used completely relaxed orbitals (with respect to the fractionally charged species), and therefore the quadratic term in Eqn. \ref{quad} could not solely arise from the quadratic dependence of classical electron-electron repulsion $E_J$, as is often thought. Residual self-repulsion is still likely a major contributor (as was noted in Ref [\onlinecite{zhang1998challenge}]), but the universality of Eqn \ref{quad} appears somewhat surprising, considering the broad spectrum of functional forms for $E_{xc}$ studied  (with varying levels of empiricism)  and orbital relaxation effects. Furthermore, HF theory (which is self-repulsion free) also yields linear $f(x)$, suggesting that the correlation contribution is quite important as well, and contributes to the overall quadratic form. It is also useful to note that HF theory with \textit{unrelaxed} orbitals satisfies Eqn \ref{linear} unlike HF with fully relaxed orbitals\cite{li2017piecewise}, indicating that orbital relaxation effects have substantial impact on delocalization error and contributes to the overall linearity of $f(x)$.}

\begin{figure}[htb!]
	\centering
	\includegraphics[width=0.5\linewidth]{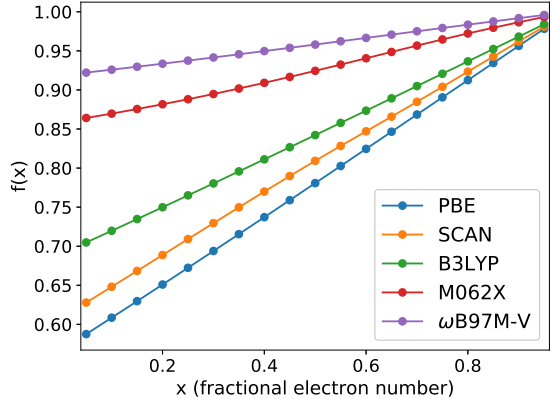}
	\caption{$f(x)$ for fractional ionization of water with representative functionals.}
	\label{fig:waterplots}
\end{figure}

Several fits of $f(x)$ for the water molecule are given in Fig \ref{fig:waterplots}, which reveal the typical extent to which linearity is observed. This linear form is rather convenient as the entirety of delocalization error is in principle encoded by the curvature $a$. \textcolor{black}{Since $E(N-1)-E(N)$ is universally positive for stable ground states (as otherwise the $N$ electron species would spontaneously eject an electron and form the $N-1$ electron species)}, $a>0$ is typical of the standard overdelocalization error predicted by semilocal functionals, while $a<0$ indicates overlocalization (as in HF theory). The intercept $b=1-a$ from the constraint $f(x=1)=1$, although in practice, small higher order terms make $b$ different by $\approx 0.01$ or less. \textcolor{black}{This is not too surprising as the higher order terms in $f(x)$ are expected to have maximum impact around $x=1$ , and is consistent with observations of deviations from typical curvature close to integer electron numbers in Ref [\onlinecite{li2017piecewise}]}. The area between the real $f(x)$ curve and the ideal $f(x)=1$ is therefore $\dfrac{|a|}{2}$ and similarly, the area between Eqns \ref{quad} and \ref{linear} is $\dfrac{|a|}{6}(E(N-1)-E(N))$. Functionals with smaller $|a|$ thus have smaller delocalization error, permitting assessment of the extent to which a functional satisfies Eqn \ref{linear}. This subsequently could allow characterization of whether DFAs are getting the right answers for the right reasons, or if there is a hidden cancellation of errors behind the accuracy of good functionals. 

\textcolor{black}{The generalized Janak's theorem \cite{janak1978proof,cohen2008fractional} furthermore states that the orbital energy for a fractionally charged species is given by 
\begin{align}
\epsilon(x)=\dfrac{dE(N-x)}{dx}=-(E(N-1)-E(N))(2ax+b)=-(E(N-1)-E(N))g(x)
\end{align}
which is linear in the fractional electron count. $g(x)$ therefore should have twice the slope of $f(x)$, which numerically appears to roughly be the case. A curvature of $a=0$ would thus trivially enforce Koopman's theorem.  We note that the figures in Ref [\onlinecite{tsuneda2010koopmans}] show orbital energies that are roughly linear with fractional electron number, although they did not discuss the overall implications of this behavior}. 

\begin{figure}[htb!]
	\centering
	\includegraphics[width=0.5\linewidth]{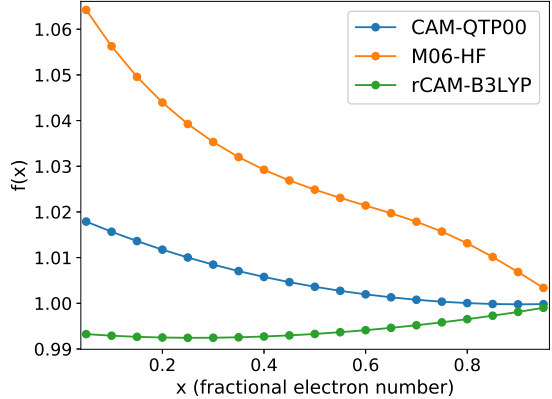}
	\caption{Non-linear $f(x)$, as defined in Eq. \ref{gen}, for fractional ionization of water.}
	\label{fig:waterdevs}
\end{figure}

It is however important to note that M06-HF\cite{m06hf}, rCAM-B3LYP\cite{rcamb3lyp} and CAM-QTP00\cite{verma2014increasing} tend to yield somewhat non-linear $f(x)$ (as can be seen in Fig \ref{fig:waterdevs}), and a few other functionals show some deviation from linearity for specific systems. Interestingly, all those cases are characterized by very low delocalization error ($f(x)$ close to $1$ despite non-linearity) relative to most functionals with linear behavior. Consequently, they yield a better approximation to the ideal $f(x)=1$ flat line than some of the linear $f(x)$ with larger slopes (which a comparison of the y axis scales in Figs \ref{fig:waterplots} and \ref{fig:waterdevs} makes evident). This indicates that these functionals were effective in eliminating the principal source of delocalization error (the quadratic term in Eqn \ref{gen}) to a substantial extent, leaving behind less significant higher order terms. This could therefore be behavior for functionals to aspire to in order to minimize delocalization error. Certain functionals like CAM-QTP01\cite{jin2016qtp} and $\omega$B97X-V\cite{wb97xv} however only show non-linear behavior for specific systems, with the non-universality suggesting that perhaps it is difficult for a single functional to be effective at addressing delocalization error over many species. Tuning functional parameters for specific systems\cite{baer2010tuned}  could offer a potential solution to this issue. Stein et.al.\cite{stein2012curvature} had in fact noted that tuning functionals to satisfy Koopman's theorem\cite{baer2010tuned,stein2009reliable} alleviates the effect of terms quadratic in fractional electron number in the delocalization error. Our work here notes that this quadratic term is the principal source of delocalization error, which perhaps explains the successful application of tuned functionals for many systems\cite{stein2012curvature,bredas,srebro2011tuned,srebro2012does,kronik2012excitation,salzner2011improved,stein2009reliable}. \textcolor{black}{It also potentially helps explain the efficacy of DFT+U\cite{cococcioni2005linear} (which uses a quadratic, subshell population based localization correction to DFT inspired by the Hubbard model), as well as various quadratic scaling corrections\cite{zheng2011improving,li2015local,bajaj2017communication,li2017localized}.}

\begin{figure}[htb!]
	\centering
	\includegraphics[width=0.5\linewidth]{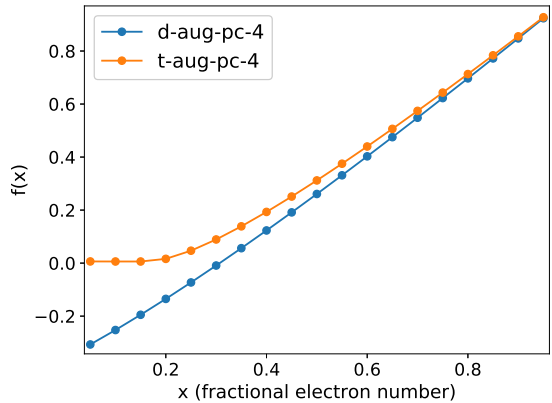}
	\caption{$f(x)$ predicted by fractional ionization of F$^-$ with the PBE functional, using the doubly augmented d-aug-pc-4 and triply augmented t-aug-pc-4 basis sets. The aug-pc-4 results are visually indistinguishible from the d-aug-pc-4 values.}
	\label{fig:augment}
\end{figure}

\textcolor{black}{Fractional ionization of the F$^-$ anion reveals some additional features that merit discussion. The $a$ values predicted by many functionals are sufficiently large that the parabolic energy curve given by Eqn \ref{quad} has an extremum between $x=0$ and $x=1$. Such an energy extremum however can only be an artifact of a finite basis set. An energy minimum at $N-x$ electrons would indicate that the $N$ electron species would energetically prefer to exist as a fractionally charged $N-x$ electron entity, with the remaining $x$ electrons pushed to infinity. This unphysical leakage of electron density has long been known to occur for calculations on anions with local functionals\cite{vydrov2007tests,jensen2010describing,peach2015fractional}, when an adequately flexible basis set is employed. Basis sets with fewer diffuse functions however have insufficient flexibility to push the extra fractional electron to infinity, and consequently spuriously predict  a higher energy for $N$ electrons relative to $N-x$ electrons due to forced localization of the extra fractional electron. Similar behavior is exhibited when an energy maximum is predicted instead (as in HF), since that would indicate that a fractional electron should refuse to bind to the $N-1$ electron species, until the resulting $N-x$ electron species will be lower in energy than the $N-1$ electron species\cite{peach2015fractional}.}

\textcolor{black}{The essentially linear $f(x)$ predicted by most functionals with the aug-pc-4 basis on F$^-$ therefore should be an artifact of the basis set which can be eliminated by addition of more diffuse functions. Fig. \ref{fig:augment} shows this in practice, with the triply augmented t-aug-pc-4 basis predicting $f(x)=0$ for PBE till a cutoff point around $x\approx 0.2$ (where aug-pc-4 results predicts an energy minima), and is afterwards linear. Similar piecewise linear behavior ($f(x)=0$ up to a cutoff, linear with non-zero slope beyond) is anticipated for other delocalizing functionals. Localizing methods like HF yield slightly different behavior---$f(x)$ is linear up to the cutoff point, and afterwards $f(x)=x^{-1}$ (since $xf(x)=1$ if the fractional electron would thereafter refuse to bind), in order to avoid spurious energy maxima. This appears to be the most significant deviation from linear $f(x)$ that occurs in the high delocalization error limit. Nonetheless, the delocalization energy error formally remains a \textit{piecewise} quadratic polynomial in the fractional electron number (with at most two pieces). 
This behavior is common for for F$^-$ using non range-separated functionals, but for CN$^-$ is an issue encountered by only LSDA and GGAs  (and that too to a very small extent as the leakage is of the order of  $10^{-2}-10^{-3}$ electrons).  Standard basis sets (e.g. with only one set of diffuse functions) furthermore appear to predict fully linear $f(x)$ due to the fractional electron being constrained to be close to the nuclei.
It is nonetheless simple to predict the location of energy `extrema' from such $f(x)$, with an extremum between $x=0$ and $x=1$ being a signature of fractional electron binding due to basis set incompleteness. The overall approach therefore serves as a test of the reliability of functional/basis pairings for any given anionic system.}

\begin{figure}[htb!]
	\centering
	\includegraphics[width=0.5\linewidth]{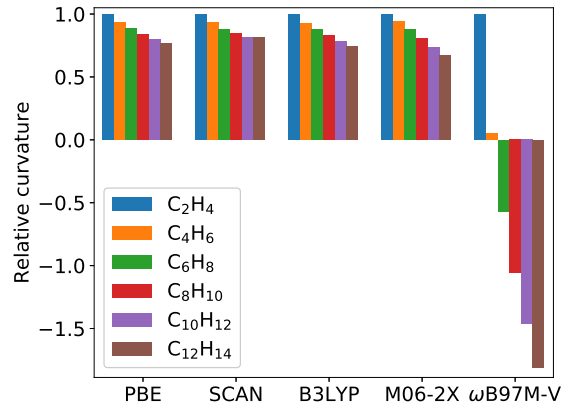}
	\caption{Changes in $a$ on increasing polyene chain length for representative functionals. $a$ for each functional has been scaled with respect the value for ethene in order to normalize the scale. }
	\label{fig:size}
\end{figure}

So far our discussion had only focused on behavior in atoms and small molecules. It is however necessary to examine whether the observed linear behavior persists for larger systems where DFT is widely used. To this end, we studied the behavior of $f(x)$ for the 47 studied functionals for fractional ionization of a series of conjugated trans polyenes ranging in size from C$_2$H$_4$ to C$_{12}$H$_{14}$, using the triple-zeta pc-2 basis. The same linear behavior was observed throughout, indicating that $f(x)$ does not appear to become nonlinear on larger length scales. Interestingly, the curvature $a$ decreased in value as the size of the polyene increased. This was not wholly unexpected for range separated hybrid (RSH) functionals, which employ more and more HF exchange on longer length scales and thus should have curvature that becomes more negative with increasing size. It is however somewhat surprising for other functionals without an explicitly defined length scale like PBE\cite{PBE} or B3LYP\cite{b3lyp}, where $a$ decreases in magnitude on increasing the size of the polyene (as can be seen in Fig \ref{fig:size}). This decrease in curvature can originate from two possible sources. The fractional charge can delocalize over a longer length for longer polyenes due to the conjugated $\pi$ system, resulting in smaller local fractional charge density at any given point and thus less resultant self-repulsion. Long polyene chains are also known to exhibit substantial strong correlation effects\cite{hachmann2006multireference,hu2015excited,lehtola2016cost} which single reference theories like HF and Kohn-Sham DFT might not be able to account for. Negative $a$ in HF originates from missing correlation energy, and it is therefore possible that the decrease in $a$ on longer polyenes originates in part from increasing strong correlation effects. HF in fact predicts more negative $a$ with increasing chain length, suggesting that strong correlation effects have a role to play in the decrease of curvature. Tests with HF and a few representative DFAs on alkanes (where strong correlation effects are expected to be minimal) however shows a decrease in the magnitude of $a$ over chain length as well for non RSH methods, indicating that reduction in self-interaction likely also plays a role in reducing curvature in polyenes.  It is unclear whether the curvature would ultimately asymptote to a finite value on increasing the chain length further, or if it will steadily decrease to a point where even a local functional would overestimate fractional charge energy due to missing strong correlation. Overall however, there is no deviation from linearity of $f(x)$ on moving to larger systems, indicating that the quadratic \textcolor{black}{polynomial} nature of delocalization error is a general feature. The performance of RSH functionals with a predefined length scale however may suffer if the functionals become too localizing on very long length scales, and some testing therefore is recommended before extensive use. 

\begin{figure}[htb!]
	\centering
	\includegraphics[width=\linewidth]{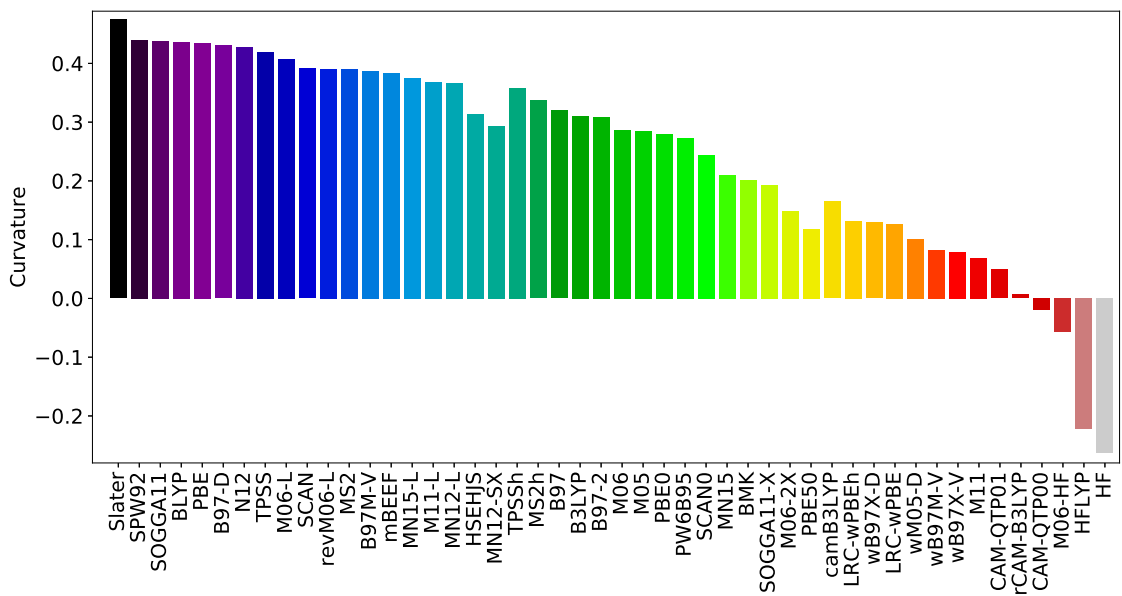}
	\caption{Curvature $a$ for water molecule for studied functionals}
	\label{fig:watercurves}
\end{figure}

Small nonlinearities aside, it is interesting to note to what extent are various functionals affected by delocalization error. This can be done via comparing the curvature $a$ for many functionals for a given species (obtained via fitting to $f(x)$), or by taking weighted averages over multiple systems to account for differences between molecules. The \textcolor{black}{relative variation in performance between functionals} observed by us however are quite small, and the spectrum of $a$ for systems like the water molecule (as shown in Fig \ref{fig:watercurves}) appears adequate for determining the extent to which various functionals eliminate delocalization error. \textcolor{black}{It is however important to note that the numerical scale of $a$ values in Fig \ref{fig:watercurves} need not be representative for all chemical problems. Involvement of anions is likely to lead to larger $a$ (as seen by us for fractional ionization of F$^{-}$ and CN$^{-}$), though the relative $a$ values should have much smaller variation from system to system}. 

Fig \ref{fig:watercurves} shows that local functionals as a rule have fairly large curvature on account of having a purely local exchange-correlation component. There is very little decrease in curvature in moving from LDA to GGA, or even from GGAs to mGGAs. The latter is particularly interesting as certain modern mGGAs like mBEEF\cite{wellendorff2014mbeef} and SCAN\cite{SCAN} exhibit hybrid-esque features like very small partial charges on dissociation of polar bonds\cite{hait2018accurate}, low polarizability errors\cite{hait2018accuratepolar}, improved band gap predictions\cite{zhang2017comparative} and Rung 4 like excited state energies\cite{tozer2018molecular}. This has been attributed to reduced delocalization error in the past\cite{hait2018accurate,hait2018accuratepolar,tozer2018molecular,zhang2017comparative}, but the comparatively high curvature of these mGGAs relative to hybrid functionals suggest that other factors that affect behavior for integer electron numbers might be responsible to a significant extent as well. For instance, the lack of fractional charges at dissociation could just as well be a consequence of overestimation of the gap between the IP of the electron donor fragment and the EA of the electron acceptor fragment. Similar factors could mask the effect of large $a$ elsewhere, but the presence of the curvature serves as a reminder about how far modern local DFAs are from the exact functional, despite substantial advances in predictive power for other applications.  

Global hybrid functionals have lower curvature than locals, due to the localizing tendency of HF partially canceling the delocalizing effect of the local exchange-correlation functionals. The value of $a$  in fact roughly decreases monotonically with the increasing HF exchange fraction, with TPSSh\cite{tpssh} (10\% HF exchange) having the largest curvature and PBE50 (50\% HF exchange) having the smallest. Functionals with 100\% HF exchange like M06-HF and HFLYP\cite{lyp} take this trend to the extreme by predicting negative $a$ like HF itself. Very interestingly, screened exchange functionals like HSEHJS\cite{hsehjs} do not appear to have substantially larger curvature than many global hybrids like PBE0\cite{pbe0}, indicating that their application to solid state problems for band gap predictions might not be too adversely affected by the lack of long range HF exchange. 

The smallest magnitudes of $a$ however are predicted by range separated hybrid (RSH) functionals. The major exception to this rule is cam-B3LYP\cite{camb3lyp}, which gives a performance somewhat between global hybrids and RSH functionals on account of it not employing 100\% HF exchange at long distances. Other RSH functionals supply some of the lowest curvatures observed, with recently parameterized functionals like M11\cite{m11}, $\omega$B97X-V\cite{wb97xv} and $\omega$B97M-V\cite{wB97MV} giving relatively very good performance. The lowest curvatures however are obtained from the CAM-QTP\cite{verma2014increasing,jin2016qtp} functionals and rCAM-B3LYP\cite{rcamb3lyp}. The performance of the former is unsurprising since they were fit to IP's of water, but the latter does very well too--perhaps as a consequence of being fitted to fractionally charged species. All three of these functionals predict low curvature for other systems as well, showing that their parameterization strategy was quite effective in addressing major contributions to delocalization error. It should however be noted that they start predicting too negative a curvature for long polyene chains like C$_{12}$H$_{14}$, and thus their performance for larger systems  (or systems with strong correlation) may not be as excellent. \textcolor{black}{In general, it might not be a bad idea to test a number of functionals to see which yields minimum $|a|$ for a given system of interest, in order to determine the suitability for applications where delocalization errors are expected to matter (like TDDFT for charge transfer states\cite{dreuw2003long,dreuw2005single})}.

The superb performance of CAM-QTP00 and rCAM-B3LYP nonetheless raises a question as to whether it is possible parameterize other functionals with low curvature that are simultaneously highly accurate for \textcolor{black}{general} applications. One evident way to achieve this would be through fitting to fractional charges. Eqn \ref{quad} appears to be the default behavior for most functionals, and it is completely specified by three parameters: the integer electron energies $E(N),E(N-1)$ and the curvature $a$ (applying the exact constraint of $b=1-a$). Current fits to IPs and EAs already employ information about $E(N)$ and $E(N-1)$. It seems straightforward to average benchmark integer electron energies to obtain $E(N-0.5)$, and add it to the database used for training and/or testing purposes. These three points are sufficient to constrain the quadratic form of Eqn \ref{quad} (which functionals naturally seem to predict) to the ideal linear connection predicted by Eqn \ref{linear} for each individual case. Additional energies for different fractional charges would therefore be superfluous if $f(x)$ was perfectly linear. Of course, the presence of higher order corrections prevent this from being strictly true, and their magnitude could increase as the quadratic term is damped out. However, the performance of functionals with the least linear $f(x)$ suggest that these higher order effects would still be small and the greater part of delocalization error could be canceled out in this fashion. Finally, it ought to also be possible to tune functionals to accurately reproduce the energy of the $E(N-0.5)$ state (which in practice would likely involve minimization of the magnitude of $2E(N-0.5)-E(N)-E(N-1)$), as an alternative to Koopman's theorem based tuning, to eliminate delocalization errors. \textcolor{black}{The advantage of the proposed method is that it is entirely energy based, and thus can be applied to circumstances where orbital identities are unclear, like Rung 5  double hybrid functionals}.

In summary, we have demonstrated the near quadratic \textcolor{black}{polynomial} dependence of the fractional electron delocalization error via strongly linear behavior of a delocalization function $f(x)$. This behavior was observed over a number of small systems, as well as moderately long polyene chains, indicating widespread occurrence in chemical systems for many functionals. The slope of this delocalization function (which corresponds to curvature of the delocalization energy error) can be used as a metric to characterize the extent to which functional-system pairs suffer from delocalization error. Furthermore, while a few functionals possess relatively non-linear $f(x)$, they are also characterized by lower delocalization error than most of the ones showing linear behavior. This suggests that strategies (whether based on tuning or fitting/selection) that seek to minimize the magnitude of the slope of $f(x)$ would consequently eliminate the leading source of delocalization error in DFAs. The generally near-linear behavior of $f(x)$ further indicates that this minimization could be adequately accomplished using just one non-integer electron data point, such as the energy of a system with $\pm 0.5$ charge. Further investigations are required to determine if these strategies can be employed to generate tuned functionals with high system-specific utility or can be generalized to train/select DFAs with broad applicability that also have decreased delocalization errors. 
\section*{Computational Methods}
All calculations were done with a development version of the Q-Chem 5.0 software\cite{QCHEM4}. All $f(x)$ calculations were done using unrestricted orbitals. The large aug-pc-4 basis\cite{jensen2001polarization,jensen2002polarization,jensen2002polarizationiii,jensen2004polarization,jensen2007polarization} was used for the species in Table \ref{atoms}, which ought to be very close to the complete basis set limit \textcolor{black}{with respect to valence and polarization functions, though not diffuse functions. The deviation from the diffuse basis set limit is likely inconsequential for the neutral to cation segments of the fractional electron curve (as indicated by trial calculations with extra diffuse functions), but it is important for the anion to neutral segments due the possibility of `leakage' of a fractional electron to infinity. More highly augmented basis (like d-aug-pcc-4 and t-aug-pc-4) were obtained from aug-pc-4, using the geometric sequence protocol described in Ref [\onlinecite{woon1994gaussian}]}. The smaller pc-2 basis was used for the polyene chains. All local xc integrals were calculated on a grid with 99 radial and 590 angular Lebedev points. Non-local VV10\cite{vv10} correlation was computed on an SG-1 grid\cite{gill1993standard}. The geometries employed were either experimental or optimized with MP2/cc-pVTZ (full listing given in Supporting Information).
\section*{Supplementary Information}
Geometries (zip) and $f(x)$ values (excel). 

\section*{Acknowledgment} 
\textcolor{black}{D.H. would like to thank Dr Narbe Mardirossian for helpful discussions}. This research was supported by the Director, Office of Science, Office of Basic Energy Sciences, of the U.S. Department of Energy under Contract No. DE-AC02-05CH11231. D.H. was also funded by a Berkeley Fellowship.  
\bibliography{references}
\end{document}